\documentclass[12pt]{article}
\usepackage{graphicx}
\usepackage{amssymb}
\usepackage{hyperref}
\def\DY{{\rm d} y\,}
\def\sfrac#1#2{\mbox{$\frac{#1}{#2}$}}

\begin{document}

\title{Laminar-turbulent patterning in wall-bounded\\
shear flows: a Galerkin model}
\author{{\Large K. Seshasayanan$^1$ and P. Manneville$^2$}\\[2ex]
$^1$Laboratoire de Physique Statistique, CNRS UMR 8550,\\ \'Ecole Normale Sup\'erieure, F-75005 Paris, France\\[1ex]
$^2$Laboratoire d'Hydrodynamique, CNRS UMR7646,\\
 \'Ecole Polytechnique, F-91128, Palaiseau, France}
 \date{ to appear in Fluid Dyn. Res.}
%\ead{}
\maketitle

\begin{abstract}

On its way to turbulence, plane Couette flow -- the flow between counter-translating parallel plates -- displays a puzzling  steady oblique laminar-turbulent pattern.
We approach this problem {\it via\/} Galerkin modelling of the Navier--Stokes equations.
The wall-normal dependence of the hydrodynamic field is treated by means of expansions on functional bases fitting the boundary conditions exactly.
This yields a set of partial differential equations for the spatiotemporal dynamics in the plane of the flow.
Truncating this set beyond lowest nontrivial order is numerically shown to produce the expected pattern, therefore improving over what was obtained at cruder effective wall-normal resolution.
Perspectives opened by the approach are discussed.
\end{abstract}

\vspace{2pc}

\maketitle
\sloppy 

\section{Context\label{S1}}

Turbulent flows display transport properties  strongly enhanced with respect to those of laminar flows, a feature that has particularly important consequences in configurations of engineering interest.
Understanding how a given laminar flow becomes turbulent or a turbulent flow decays to laminar is therefore of great interest, both conceptual and practical.
In this respect, the case of wall-bounded flows is of utmost concern since the transition can be direct, without the intermediate steps observed, e.g. in free shear flows~(Huerre \& Rossi 1998, Manneville 2015).
This direct transition is a result of the local stability  of laminar flow in competition with nontrivial solutions to the Navier--Stokes equation (NSE for short) in the range of control parameter where the transition effectively takes place.
As analysed by Waleffe~(1997), the mechanism by which such nontrivial solutions exist, the self-sustainment process (SSP), is now thought to be well understood, but the laminar-turbulent coexistence still raises important questions.
In Hagen--Poiseuille flow (HPF), the flow under pressure gradient through a circular pipe, the transition takes place when turbulent {\it puffs\/}, the nontrivial states alluded to above, split and propagate turbulence before they have time to decay, a scenario well-reproduced by a reaction-diffusion model introduced by Barkley~(2011a).
In its own transitional range, plane Couette flow (PCF), the simple shear flow developing between counter-translating plates, experiences laminar-turbulent coexistence in the form of steady oblique bands (Prigent {\it et al.} 2002, Duget {\it et al.\/} 2010, Barkley \& Tuckerman 2005).
The Reynolds number, the relevant control parameter, is here defined as $R=Uh/{\nu}$ where $U$ is the speed of the plates, $h$ the half gap width between them, and $\nu$ the kinematic viscosity of the fluid.
The {\it bands\/} are observed for $R_{\rm g} < R < R_{\rm t}$.
Below the global stability threshold $R_{\rm g}$ turbulence is only transient, in the form of finite-lifetime {\it spots\/}, and the laminar base flow is always recovered after the spots have decayed.
Beyond the upper threshold $R_{\rm t}$ turbulence is essentially {\it featureless\/}, i.e. uniform.
A model, also of reaction-diffusion type, was proposed by one of us~(Manneville 2012) to account for this pattern formation, in which a Turing mechanism was proposed to be responsible for the bands when $R$ is decreased below $R_{\rm t}$.
Such explicative models are analogical in essence.
Trying to support them  directly from a reliable simplification of the primitive problem in order find out the physical mechanisms behind laminar-turbulent coexistence is the actual purpose of the work presented here. 

In the transitional range,  the nontrivial solutions appear to be strongly coherent at the scale of the distance to the wall, pipe diameter~(Hof {\it et al.} 2004) or gap between plates~(Bottin {\it et al.} 1998).
This feature can be incorporated in the sought-for models using Galerkin methods that project the solutions and their dynamics on well-chosen functional bases~(Finlayson 1972).
Such a method was used by Waleffe to build a dynamical system implementing the SSP for PCF with stress-free boundary conditions directly from the NSE~(Waleffe 1997).
His model is a system of ordinary differential equations governing the amplitude of velocity components involved in the SSP upon assuming full coherence at the scale of a Minimal Flow Unit~(Jim\'enez \& Moin 1991), with size of the order of the distance between the walls.
Though valuable to discuss the SSP, this assumption is not appropriate to study the extended systems of interest, long pipes or wide channels.

The method can however be adapted to such cases for which the variables have to be field amplitudes governed by partial differential equations still involving spatial coordinates rather than scalars functions of time satisfying ordinary differential systems.
Basically, coherence in the flow is taken into account by projecting the hydrodynamic variables onto a limited number of  wall-normal modes with corresponding amplitudes depending on the remaining coordinates, axial for pipe flow, in-plane for PCF.
In spirit, this modelling  approach can be considered as the analytical implementation of a direct numerical simulation (DNS) scheme in which the wall-normal spectral resolution would be varied in a controlled fashion.
When practiced in a strictly numerical context (Manneville \& Rolland 2011), this strategy showed that laminar-turbulent bands in transitional PCF are remarkably robust since they are preserved upon drastically reducing the number of Chebyshev polynomial used to represent the wall-normal dependence of the flow.
As the resolution was decreased, the quantitative price to be paid was a progressive narrowing of the  Reynolds number range where the bands were observed, accompanied by a downwards shift of that range explained by an aborted energy transfer towards small scales, whereas the main properties of the pattern were preserved qualitatively speaking.

Performing more work ``by hand'' would yield equations that would be more cumbersome than the NSE but would encode wall-normal coherence at the moderate Reynolds numbers of interest in a crucial way.
Eliminating supposedly less relevant degrees of freedom, we can hope for better physical understanding and for higher numerical efficiency since the physical dimension of the problem would then be reduced from three to two.
The limit would of course be that the obtained equations be still manageable.
Here, we present an extension of a previous Galerkin model by Lagha and one of us (Lagha \& Manneville 2007a) that pursues this agenda.
In that study, the model was truncated at lowest significant order and retained only three fields.
It displayed most of the expected qualitative properties except for the presence of any organised laminar-turbulent coexistence in wide domains~(Manneville 2009).
Taking into consideration the previously mentioned simulation results at reduced resolution (Manneville \& Rolland 2011) we surmised that this deficiency would be corrected by truncating the expansion at a higher level.
With four more fields, numerical simulations display the expected patterns as described in section~\ref{S3}, therefore validating the approach sketched in section~\ref{S2}.
Perspectives opened by this approach will be discussed in section~\ref{S4}.
The explicit expression of the model is given in~\ref{A}.

\section{Model\label{S2}}

The derivation follows previous work in (Lagha \& Manneville 2007a) with the difference that, in order to avoid difficulties in the treatment of the pressure field, the NSE governing the departure from laminar flow is now written in a velocity-vorticity formulation as described in~(Schmid \& Henningson 2001), p.155ff, i.e. the (nonlinear) Orr--Sommerfeld equation for the wall-normal velocity component $v$:
\begin{equation}
(\partial_t + u_{\rm b} \partial_x) \nabla^2 v -u''_{\rm b}\partial_x v + \mathcal N_v = \nu \nabla^4 v\,, \label{eq_vv}
\end{equation}
and the Squire equation for the wall-normal vorticity component $\zeta=\partial_z u-\partial_x w$, where $u$ and $w$ are the streamwise $(x)$ and spanwise $(z)$ velocity components, respectively:
\begin{equation}
(\partial_t + u_{\rm b} \partial_x) \zeta + u'_{\rm b}\partial_z v +\mathcal N_\zeta = \nu \nabla^2\zeta\,. \label{eq_zeta}
\end{equation}
In these general equations the base flow is $\mathbf v_{\rm b}=u_{\rm b}(y) \mathbf e_x$.
When dealing with PCF,  using the half-gap width $h$ as length unit and $h/U$ as time unit, with $U$ the speed of the plates driving the flow at $y=\pm1$ we have $u_{\rm b}(y)\equiv y$.
In that system of  units  the Reynolds number $R$ is just numerically equal to $1/\nu$, i.e. the inverse of the kinematic viscosity of the fluid. Primes denote the differentiation with respect to the wall-normal coordinate $(y)$, hence $ u'_{\rm b}\equiv1$ and $u''_{\rm b}\equiv0$.
The nonlinear terms $\mathcal N_v$ and $\mathcal N_\zeta$ are complicated, formally quadratic, expressions of the velocity components and their derivatives that can be found in~(Schmid \& Henningson 2001).
It will turn out convenient to use a poloidal-toroidal decomposition of the hydrodynamic fields by introducing  a velocity potential $\phi$ and  a stream function $\psi$ such that:
\begin{eqnarray}
\label{E-v}
&& v=-\Delta\phi\,,\label{eq:phi}\\
\label{E-uw}
&& u=-\partial_z \psi+\partial_{xy} \phi,\ w=\partial_x \psi+\partial_{zy}\phi\,,\ \zeta=-\Delta\psi\,,\label{eq:psi}
\end{eqnarray}
$\Delta$ denoting the in-plane Laplacian $\partial_{xx}+\partial_{zz}$.

\begin{figure}
\begin{center}
\includegraphics[width=0.4\textwidth]{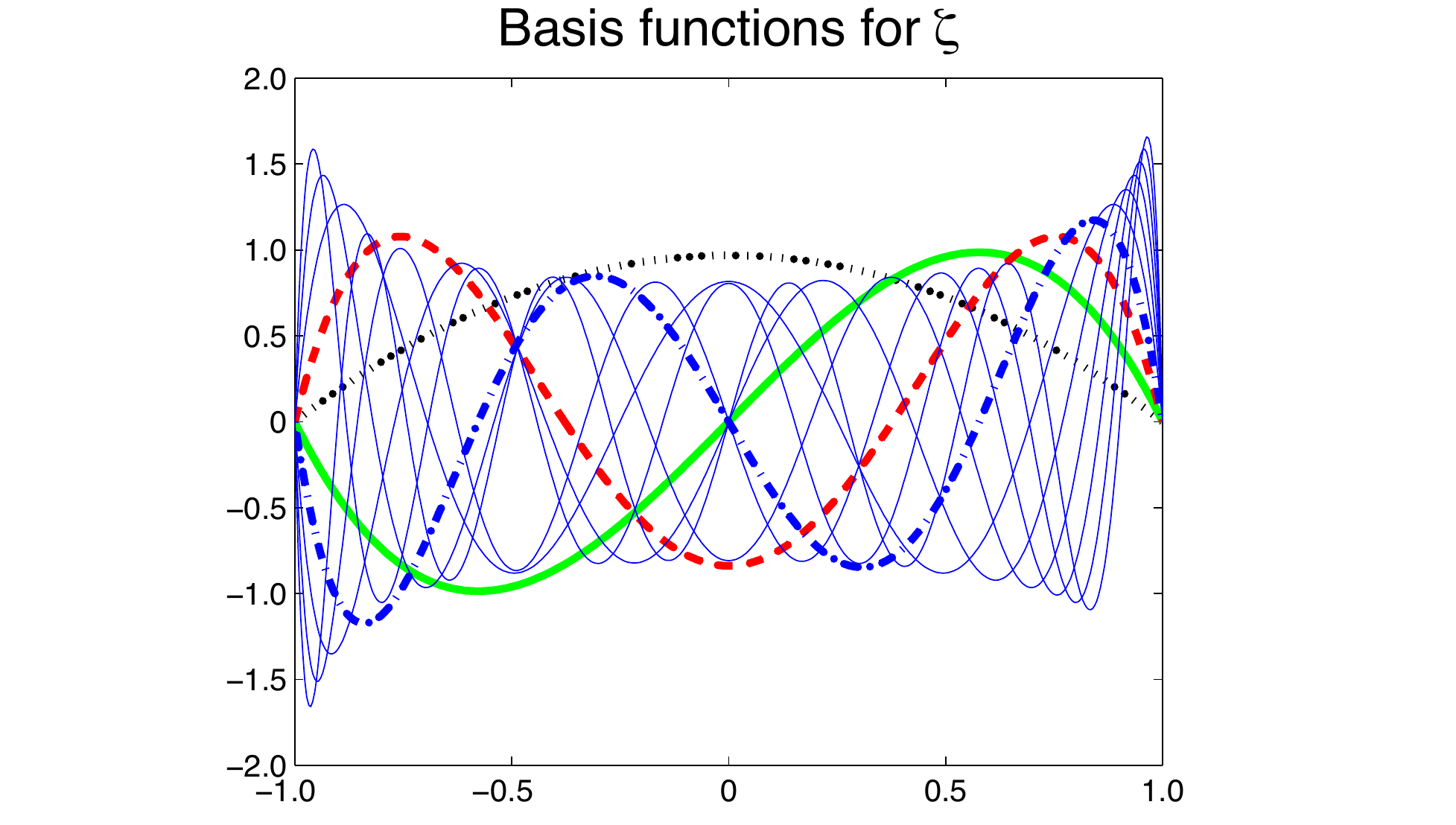}\hskip2em
\includegraphics[width=0.4\textwidth]{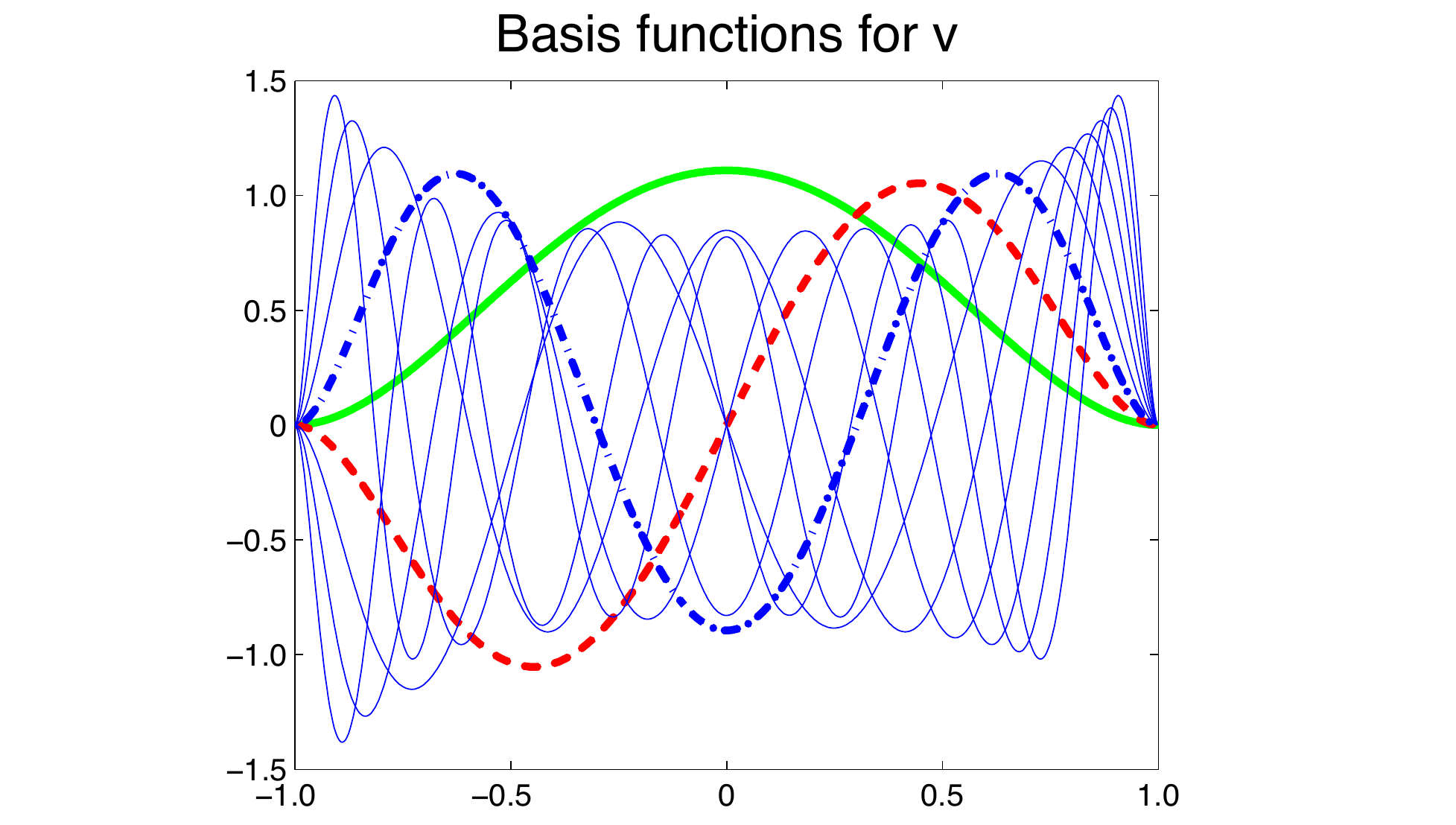}
\caption{
Basis functions for the wall-normal vorticity $\zeta$ (left) and the wall-normal velocity $v$. Functions used in our model are displayed with thick lines; $f_0$: black, dotted; ($f_1,g_1$): green, continuous; ($f_2,g_2$): red, dashed; ($f_3,g_3$): blue, dash-dotted. Higher-order functions are shown with thin lines. The work in~(Lagha \& Manneville 2007a,b) made use of $\{ f_0, f_1, g_1\}$ only.\label{fig:functions}}
\end{center}
\end{figure}

The Galerkin approach used in (Lagha \& Manneville 2007a) separates the in-plane space dependence of the hydrodynamic field from its wall-normal dependence by expanding it onto a polynomial basis in $y$, yielding amplitudes functions of  $x,z$ and time $t$.
The no-slip boundary conditions to be fulfilled read $u = v= w = \zeta = \psi = \phi = 0$ at $y=\pm1$ and, from the continuity condition $\partial_x u+\partial_y v+ \partial_z w=0$,  $\partial_y v=\partial_y \phi =0$ at $y=\pm1$.
The basis functions are chosen so as to fulfil these boundary conditions exactly.
The functions chosen for $u,w,\zeta$, and $\psi $ are in the form $f_i(y)=(1-y^2) R_i(y)$, $i=0,\dots,i_{\rm max}$; the $R_i$ are polynomials of degree $i$, and $i_{\rm max}$ is the truncation order.
For $v$ and $\phi$ the functions are taken as $g_i(i)=(1-y^2)^2 S_i(y)$, $i=1,\dots,i_{\rm max}$; the $S_i$ are polynomials of degree $i-1$ for consistency with the continuity condition at given $i_{\rm max}$.
The bases $\{ f_i\}$ and $\{g_i\}$ are separately made orthonormal {\it via\/} a standard Gram--Schmidt procedure using the canonical scalar product $\langle r | s \rangle=\frac12 \int_{-1}^{+1} r(y) s(y) {\rm d}y$.
Basis functions are shown in figure~\ref{fig:functions}
from which it is clearly understood how ({\it i\/}) the resolution close to the plates is improved by increasing the truncation order, and  ({\it ii\/}) the profiles chosen for $v$ incorporate the boundary condition $\partial_y v(x,\pm1,z,t)=0$.
The analytic expressions of basis functions up to $i_{\rm max}=5$ are given in~\ref{Af}.
As pointed out by Rolland (2012) in Appendix  B of his PhD thesis, the chosen basis $\{f_i,g_i\}$ is related to Jacobi polynomials of alternate possible use in standard spectral methods for the NSE (Canuto {\it et al}. 2007).

According to the standard Galerkin procedure~(Finlayson 1972), the expansions $\{v,\phi\}=\sum_i \{V_i,\Phi_i\}(x,z,t) g_i(y)$ and $\{u,w,\zeta,\psi\} =\sum_i \{U_i,W_i,Z_i,\Psi_i\}(x,z,t) f_i(y)$ are inserted in the equations which are then projected onto the relevant bases, Eq.~\ref{eq_vv} for $v$  onto $\{ g_i\}$, and Eq.~\ref{eq_zeta} for $\zeta$ onto $\{f_i\}$.
The concrete derivation is straightforward and  can be automated once the order of truncation $i_{\rm max}$ has been fixed.
The formal expression of the model reads:
\begin{eqnarray}
\left\{\left(\mathbb I \Delta + \mathbb A\right) \partial_t + \left( \bar\mathbb I \Delta + \bar\mathbb A \right ) \partial_x- \nu \left(\mathbb I \Delta^2 + 2 \mathbb A \Delta + \mathbb P \right)\right\} \Delta \Phi &=& \mathbb N^{(V)}\,,\label{M1}\\
\left\{\left(\mathbb I \partial_t + \mathbb B \partial_x\right) - \nu \left( \mathbb I \Delta + \bar \mathbb P \right) \right\} \Delta \Psi  + \bar \mathbb B \partial_z \Delta \Phi &=& \mathbb N^{(Z)}\,.\label{M2}
\end{eqnarray}
In these expressions $\Phi$ and $\Psi$ respectively stand for arrays $\{\Phi_1,\dots,\Phi_{i_{\rm max}})\}^{\rm t}$ and $\{\Psi_0,\dots,\Psi_{i_{\rm max}})\}^{\rm t}$, superscript `t' denoting transposition. $\mathbb I$ is the identity matrix of order $i_{\rm max}$ in Eq.~\ref{M1} and $i_{\rm max}+1$ in Eq.~\ref{M2}. $\bar\mathbb I$ is a square but non-diagonal matrix of order $i_{\rm max}+1$, playing a role similar to $\mathbb I$.
All other matrices, $\mathbb A$, $\bar \mathbb A$, $\mathbb B$,  $\bar \mathbb B$, $\mathbb P$, and  $\bar \mathbb P$ are either square or rectangular, with coefficients straightforwardly obtained by integration over $[-1,1]$ of the appropriate products of $f_i$, $g_j$ and their derivatives.
About one half of the possible combinations cancel due to parity considerations.
Remarkably enough, matrices $\mathbb A$, $\mathbb P$ and $\bar \mathbb P$ are diagonally dominant [$(i,i) \gg (i,i+k)$] and the absolute value of the diagonal terms increases rapidly with the position of the coefficient [$(i,i)\ll (i+1,i+1)$], which suggests possible simplifications in the equations governing the dynamics of the field amplitudes.
Finally, $\mathbb N^{(V)}$ and $ \mathbb N^{(Z)}$ are complicated, formally quadratic expressions of the ($U_i,V_i,W_i$)'s that have to be derived from the $(\Phi_i,\Psi_i)$'s introduced upon elimination of the pressure.
Their explicit expressions are given in~\ref{Ag}.

Equations~(\ref{M1}--\ref{M2}) only involve $\Delta \Phi$ and $\Delta \Psi$, which implies that some care is needed when dealing with spatially averaged terms corresponding to Fourier modes at $(k_x,k_z)=(0,0)$.
This is the price to pay for having used the velocity-vorticity formulation that avoid the explicit treatment of the pressure field (Schmid \& Henningson 2001).
It is then convenient to identify the uniform contributions to $u$ and $w$ explicitly by assuming:
$$
u=\bar u -\partial_z \tilde\psi+\partial_{xy} \phi,\quad w=\bar w+\partial_x\tilde\psi+\partial_{zy}\phi\,,
$$
with $\bar u$ and $\bar w$ still function of $y$ and $t$ but independent of $x$ and $z$, while $\tilde\psi$ refers to the $(x,z)$-varying part of $\psi$.
Notations being unambiguous, the tilde will be dropped in the following.
On general grounds,  the mean flow components $\bar u$ and $\bar w$ are governed by
\begin{equation}
\label{eq-ns1b}\partial_t \bar u - \nu (\bar u)''=-(\overline{uv})' \,,\qquad
\partial_t \bar w-\nu(\bar w)''=-(\overline{wv})'\,,\label{eq:uw}
\end{equation}
where the overline means averaging over the in-plane coordinates.
In the model, this is treated by expanding  $\bar u$ and $\bar w$ onto basis $\{f_i\}$.
From the continuity equation we get:
\begin{equation}
U = \overline U - \partial_z \Psi +  \mathbb C\, \partial_x\Phi\,,\quad W = \overline W + \partial_x \Psi + \mathbb C\, \partial_z\Phi\label{eq:UW}
\end{equation}
where $\overline U$ and $\overline W$ stand for arrays $\{\overline U_0,\dots \overline U_{i_{\rm max}}\}^{\rm t}$ and $\{\overline W_0,\dots \overline W_{i_{\rm max}}\}^{\rm t}$ while matrix $\mathbb C$ arises from the projection of $\partial_y v$ onto the basis used to expand $u$ and $w$ in the continuity equation.
Upon projection, equations~\ref{eq:uw} read:
\begin{equation}
\mathbb I \partial_t \overline U - \nu \bar \mathbb P \overline U =  \mathbb N_0^{(x)},
\quad
\mathbb I \partial_t\overline W - \nu \bar \mathbb P \overline W =  \mathbb N_0^{(z)}
\label{eq:baruw}
\end{equation}
where $ \mathbb N_0^{(x)}$ and $ \mathbb N_0^{(z)}$ are the projections of the $(x,z)$ spatially averaged nonlinear terms in equations~\ref{eq:uw}.
The model is now complete and ready for use.

\section{Validation\label{S3}}
By construction, the model possesses all the properties requested to account for the transitional regime of PCF: it can be checked that laminar flow is linearly stable for all Reynolds numbers, despite transient energy growth linked to lift-up, and that its nonlinearities redistribute but conserve the kinetic energy contained in finite amplitude perturbations.
A numerical solver was developed in order to examine whether bands can be recovered beyond lowest nontrivial truncation order $i_{\rm max} =1$.
Wanting to add higher modes of both parities, we chose $i_{\rm max}=3$, i.e. 7 fields: $\Psi_i$, $i=(0:3)$, and $\Phi_i$, $i=(1:3)$.
In-plane space dependence was handled using a Fourier pseudo-spectral scheme that gets rid of aliasing {\it via\/} the usual $3/2$ rule~(Canuto {\it et al.\/} 2007), i.e.  in directions $\{x,z\}$, the numbers of evolving modes are $N_{\{x,z\}}$ and nonlinear terms are evaluated {\it via\/} back-and-fro FFTs with solutions reconstructed on $\frac32 N_{\{x,z\}}$ points.
Time marching was treated by formally rewriting the initial problem $\partial_t X = \mathcal L X + \mathcal N(X)$ as $\partial_t[\exp(-t\mathcal L) X]=\exp(-t\mathcal L) \mathcal N(X)$ and solving the new problem using a Runge--Kutta scheme of order~4.

\begin{figure}
\begin{center}
\includegraphics[height=0.32\textwidth]{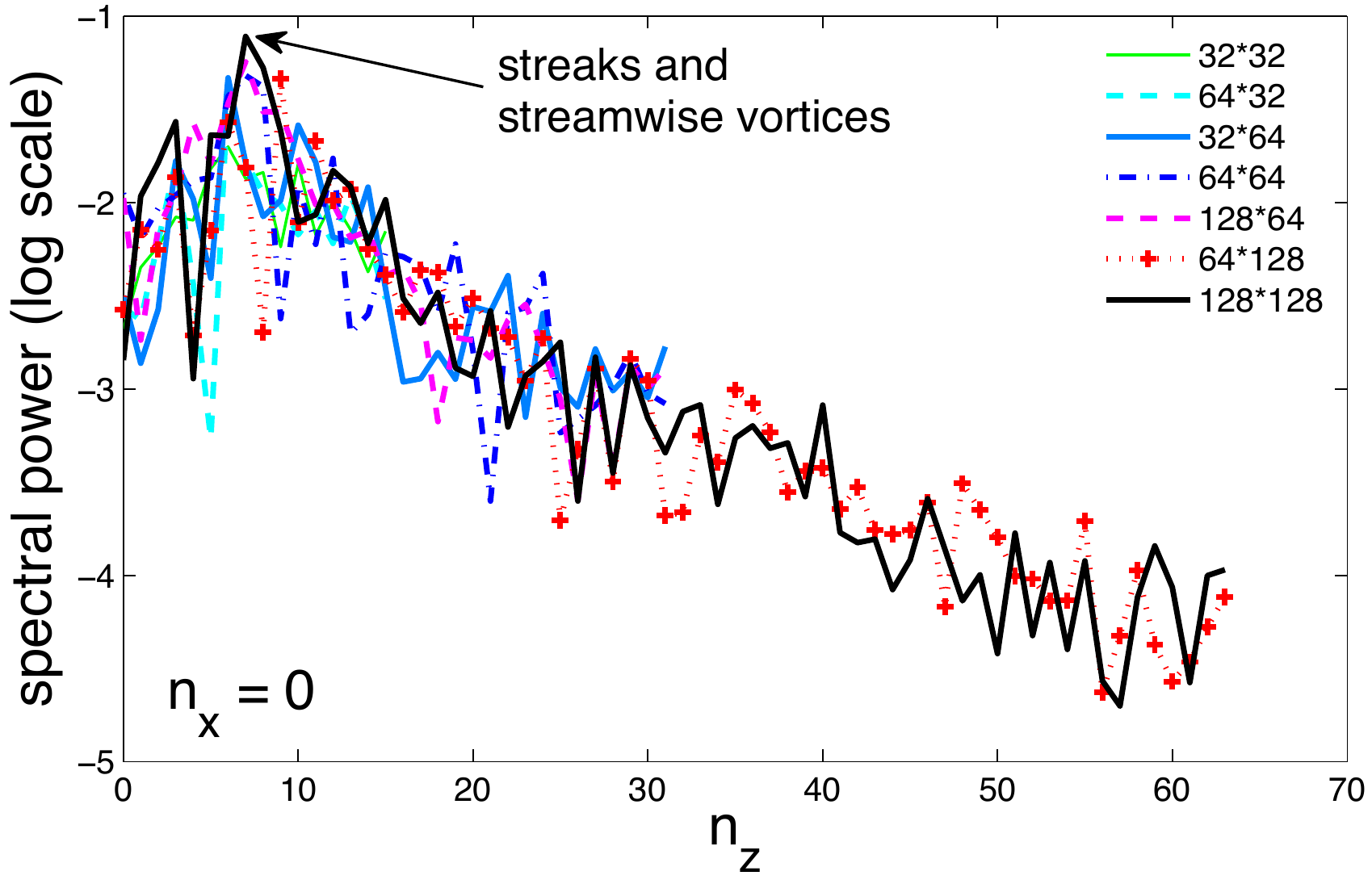}
\includegraphics[height=0.32\textwidth]{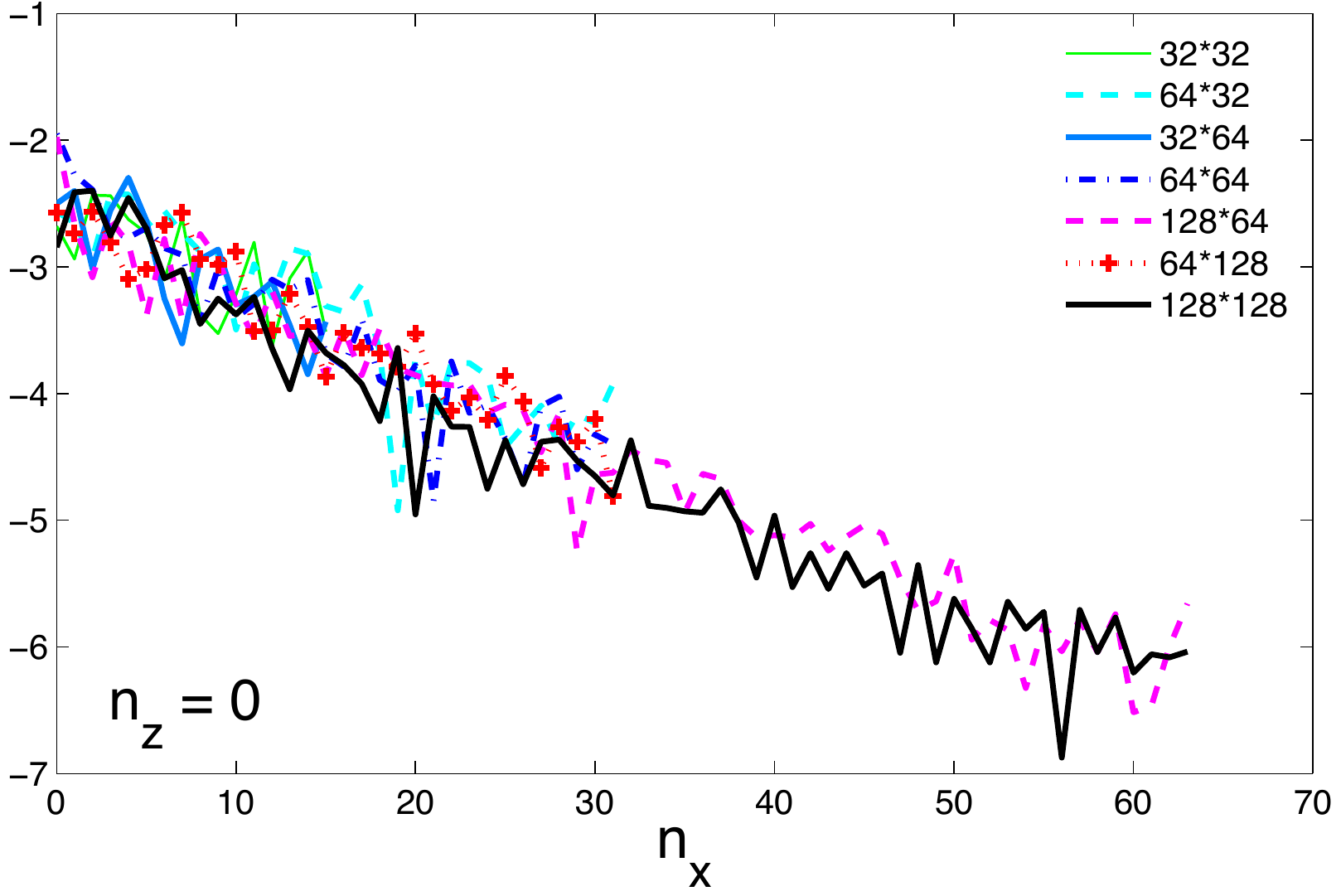}
\end{center}
\caption {Fourier spectral power of the streamwise perturbation velocity component $u$ in the mid-plane $y=0$ as functions of wave-numbers $n_z$ for  $n_x=0$ (left) and $n_x$ for $n_z=0$ (right). The corresponding Fourier wave-vectors read $k_{\{x,z\}}=2\pi n_{\{x,z\}}/L_{\{x,z\}}$, where $L_{\{x,z\}}$ are the streamwise ($x$) and spanwise ($z$) dimensions of the computational domain. The curves correspond to the different resolutions studied, in the form $N_x * N_z$, where $N_{\{x,z\}}$ are the maximum running wave-numbers.}
\label{Spectra}
\end{figure}

In parallel to the study in~(Manneville \& Rolland 2011) dealing with DNS at reduced wall-normal resolution, a first numerical experiment was devoted to the recovery of the featureless turbulent state belonging to the nontrivial branch at high $R$ in a domain of size $L_x\times L_z=32 \times 32$, aiming at an optimisation of the in-plane numerical resolution in view of the reliable simulation of domains as wide as possible at the cheapest possible computational cost.
Results shown in figure~\ref{Spectra}
display the power spectra of the streamwise component of the  perturbation velocity $u$ as a function of wave-numbers $n_z$ for $n_x=0$ (left) and of $n_x$ for $n_z=0$ (right).
Normalisation by the total number of modes $N_xN_z$ makes the curves corresponding to the different resolutions lie on top of each other.
In the left panel, the peak generated by the spanwise statistical periodicity of streaks and streamwise vortices is clearly identified for all the resolutions considered but more pronounced for $N_x\times N_z=128 \times 128$ than for $32\times 32$.
This corresponds to $N_{x,z}/L_{x,z}=4$ and 1, with effective space steps $\delta_{x,z}=0.25$ or $1$, respectively, to be compared to the period of the streaks $\lambda_z\sim L_z/n_{\rm str}$ with $n_{\rm str} \approx 7$, hence about $\lambda_z\sim 4.6$ in reasonable agreement with known results.
As seen in the right panel, the streamwise correlations decrease in a monotonic way as expected from the discussion in~(Philip \& Manneville 2011) where size effects on the temporal vs. spatiotemporal character of the dynamics was scrutinised.
The subsequent study is restricted to the configurations mentioned in the table below, with the resolutions found acceptable from the previous experiment:   
\begin{table}[h]
\centering
\begin{tabular}{c|c|c|c|}
$L_x$&$L_z$&$N_x/L_x$&$N_z/L_z$ \\
\hline
$108$&$48$&$2$&$4$ \\
$128$&$84$&$2$&$4$ \\
$680$&$340$&$1$&$1$ \\
%\hline
\end{tabular}\end{table}

\begin{figure}
\begin{center}
\includegraphics[width=0.48\textwidth]{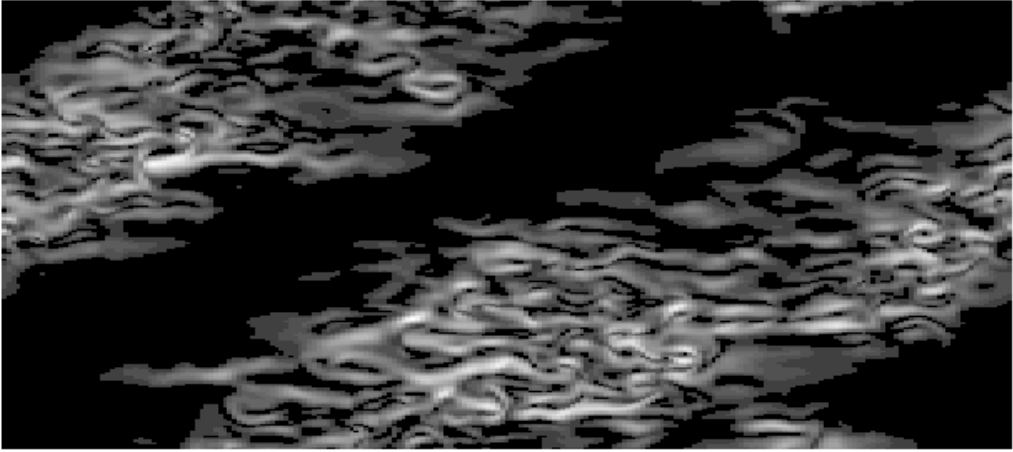}\hfill
\includegraphics[width=0.48\textwidth]{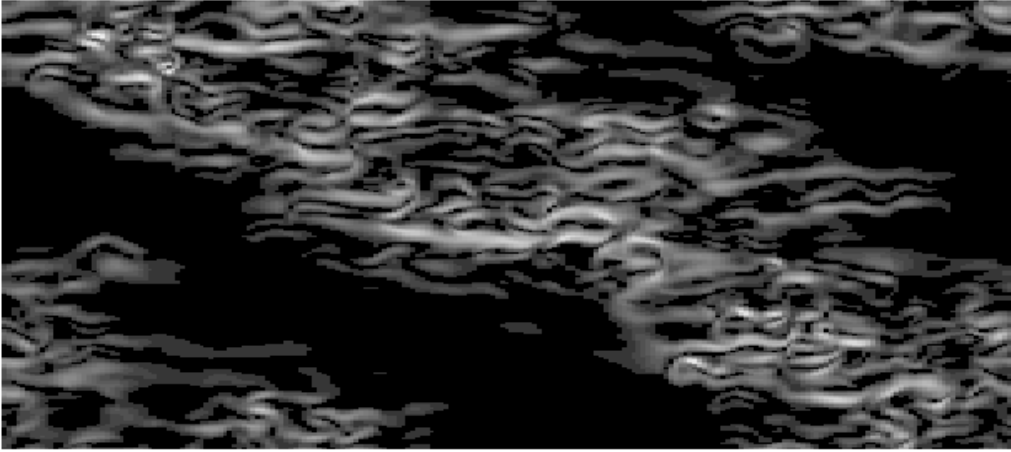}\\[2ex]
\includegraphics[width=0.95\textwidth] {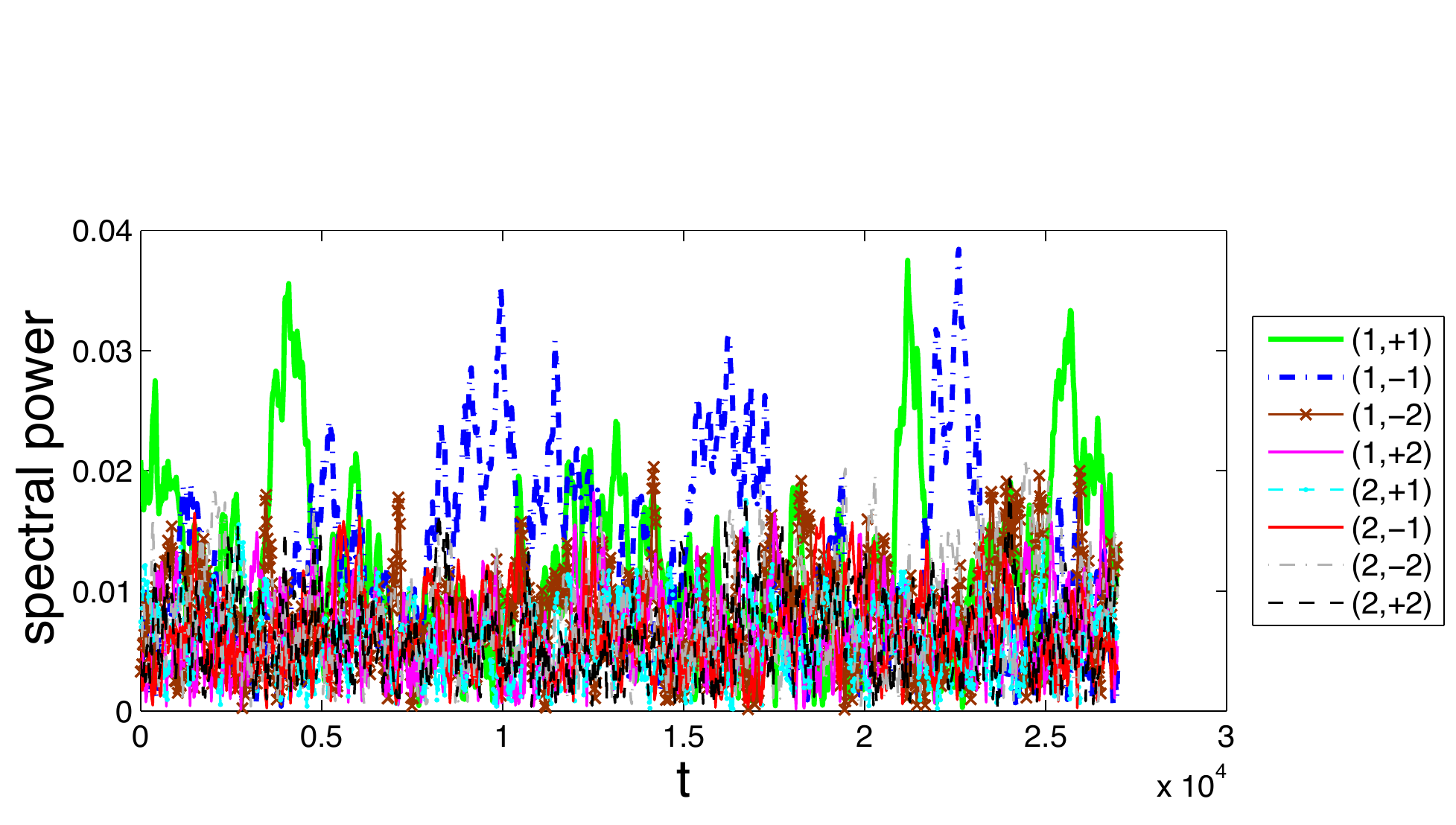}
\end{center}
\caption {Patterning in a domain of size $108\times 48$ at $R = 151$. Top: the two different orientations $(1,+1)$ (left) and $(1,-1)$ (right) with perturbation energy field averaged over the gap in grey levels, black = laminar, white = largest local energy.
Bottom: Orientation fluctuations evidenced by the spectral power in modes with wave-numbers $(n_x, n_z)$, $n_x=1,2$, $n_z=\pm1,\pm2$.}
\label{108x48}
\end{figure}

Our main result is that, in all cases, steady oblique patterns of alternately laminar and turbulent domains were observed in a limited range of Reynolds numbers, between $R_{\rm g} \approx 150$ and $R_{\rm t} \approx 159$.
Figure \ref{108x48} (top) for $R=151$ illustrates the two different possible orientations of a single band pattern in the domain $108 \times 48$.
Orientation fluctuations are known to exist in DNSs at such an intermediate size.
They are also present in the model as seen in the bottom panel showing the alternative dominance of modes $(1,+1)$ and $(1,-1)$, while the other modes are less intense.
Here fluctuations seem more important than in DNSs, with briefer episodes of well-formed pattern and a much smaller signal-to-noise ratio; compare with Fig.~3 of~(Rolland \& Manneville 2011).
\begin{figure}
\includegraphics[height=0.32\textwidth]{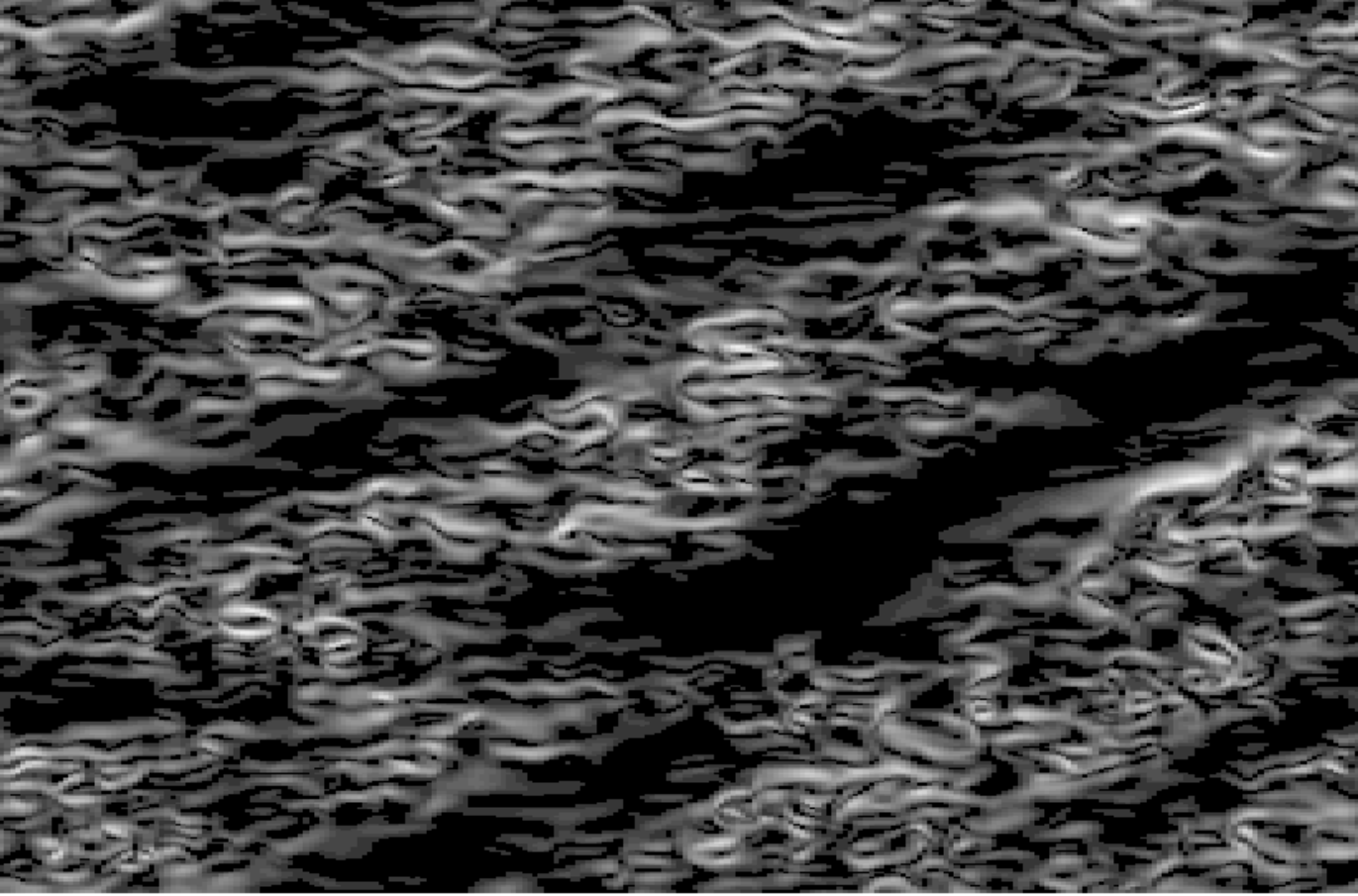}\hfill
\includegraphics[height=0.33\textwidth]{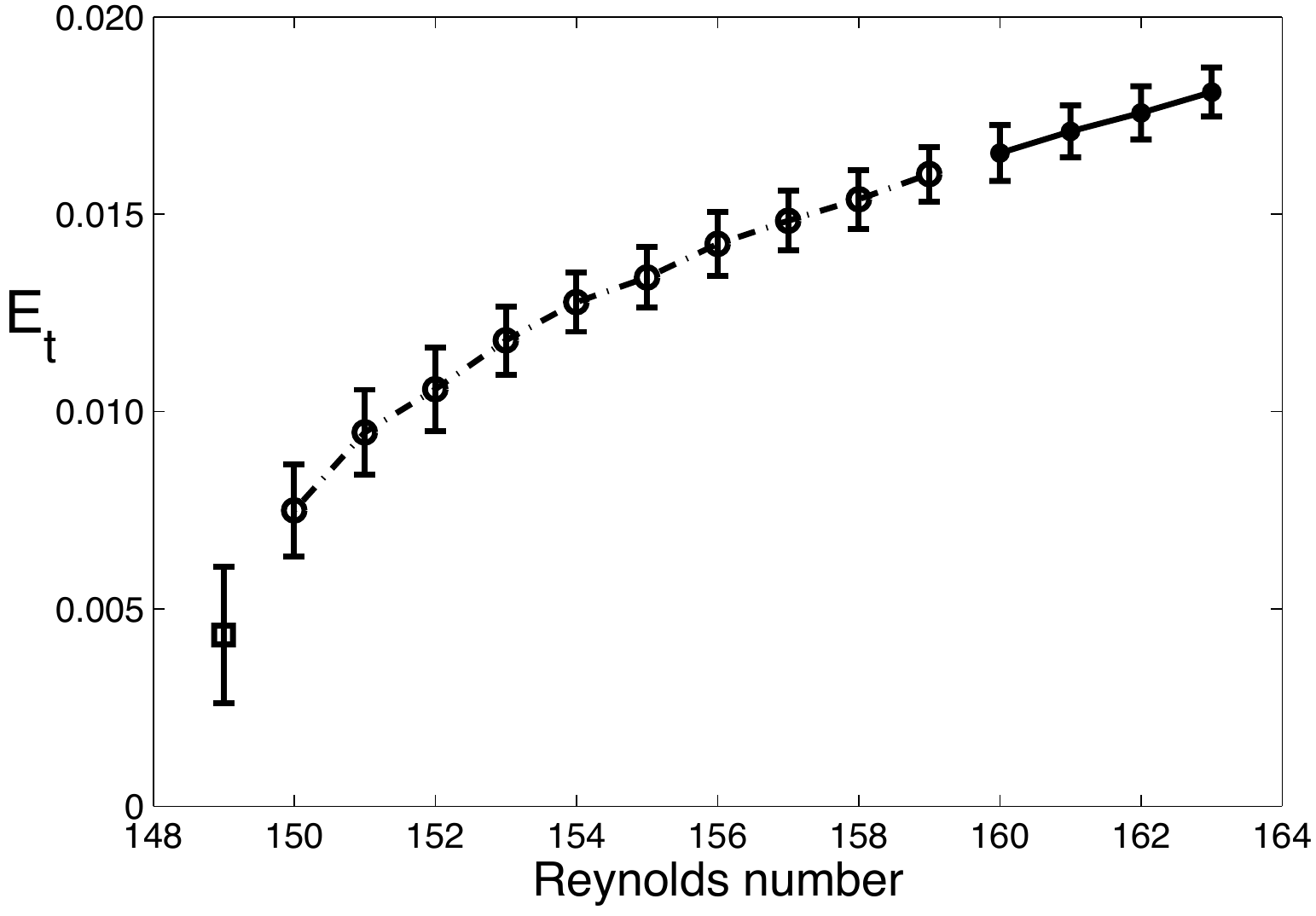}
\caption{Patterning in a domain of size $128\times 84$. Left: Patterns at $R = 154$. Right: Bifurcation diagram (distance to laminar flow as a function of Reynolds number, see text).}
\label{Bigger}
\end{figure}

Larger domains can accommodate more bands as seen in figure~\ref{Bigger} (left) for a $128\times84$ domain.
The average of the turbulent energy over the volume $V$ of the domain, $E_{\rm t} = \frac{1}{V}\int_V\frac12(u^2+v^2+w^2)\, {\rm d}x\,{\rm d}y\, {\rm d}z$, has been measured through the whole transitional range.
The bifurcation diagram displayed in figure~\ref{Bigger} (right) is again as expected, however the occurrence of large-scale laminar-turbulent coexistence in the form of oblique bands, easily detected visually and permitting the identification of $R_{\rm g}\approx150$ and $R_{\rm t}\approx159$, leaves a weaker signature on the variation of $E_{\rm t}$ with $R$ than in DNSs for which a marked break at $R_{\rm t}$ and a linear decrease below were observed.
Here the smoother variation of $E_{\rm t}(R)$ and the absence of clear-cut change at $R_{\rm t}$ between the band regime (open circles)  and uniform turbulence (filled circles) are presumably again a direct consequence of the higher level of fluctuations.
Below $R=150$, turbulence is only transient but a mean energy, roughly constant before the decay stage, can still be measured (open square).
In principle $R_{\rm g}$ should be located using a statistical study in line with the approach in terms of chaotic transients, like in (Lagha \& Manneville, 2007a).
Its detection {\it via\/} a single experiment where $R$ was progressively decreased by small steps has been judged sufficient for the present purpose.

\begin{figure}
\begin{center}
\includegraphics[width=0.7\textwidth]{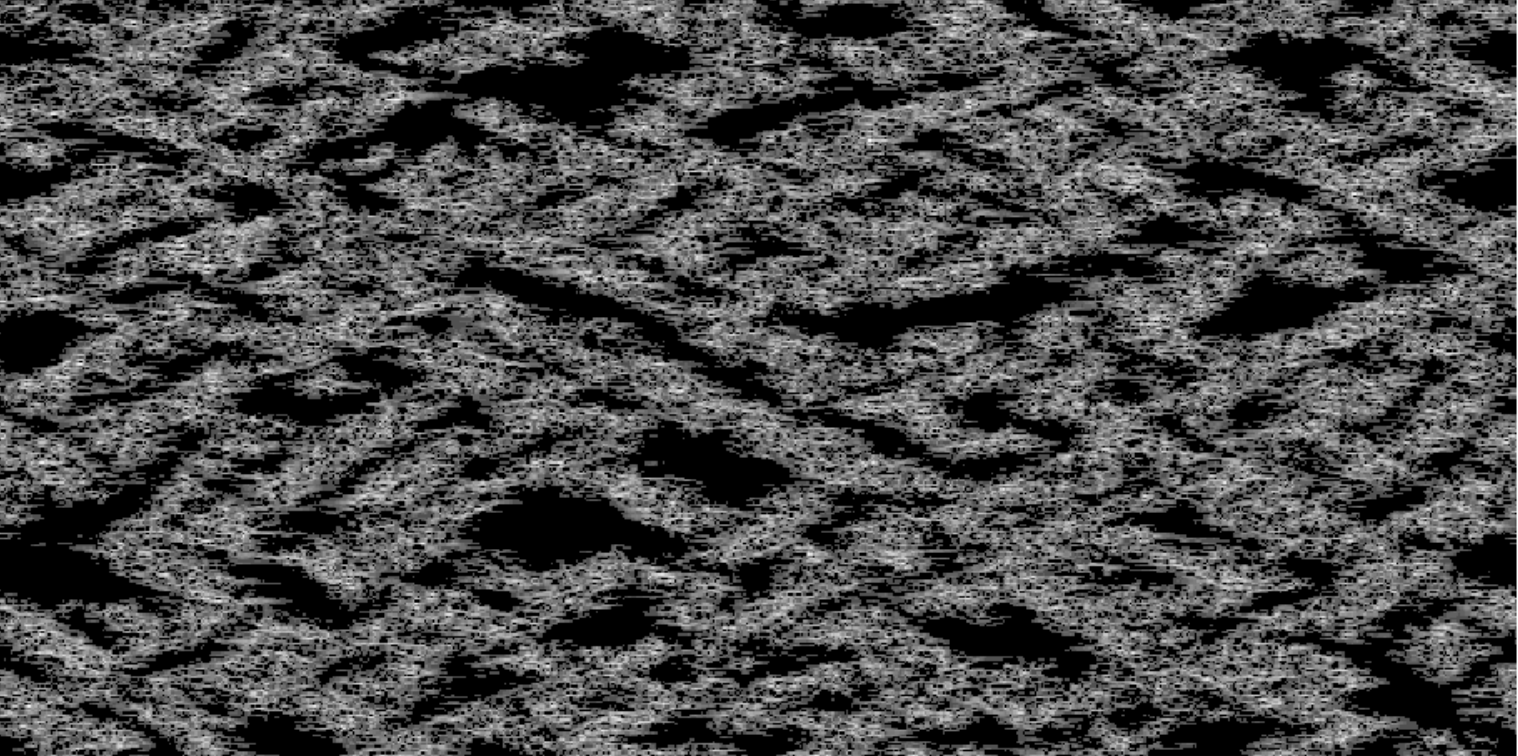}
\end{center}
\caption{Snapshot of the solution in a domain of size $680\times340$ at $R=151$. To reach such a size, the in-plane resolution has been lowered to $N_x=L_x$, $N_z=L_z$, without destroying the pattern.\label{680x340}}
\end{figure}

Finally, in a very wide domain $680\times340$ of size comparable to that of the largest experimental setups~(Prigent {\it et al.} 2002), patterns with many wavelengths were obtained.
Comparing figure~\ref{680x340}
and figure~8 in~(Manneville \& Rolland 2011) one note that the model generates outputs quite similar to what is obtained in DNSs at reduced wall-normal resolution, itself representing the experimental situation reasonably well.

Whereas at a qualitative level the model is fully satisfactory, we have however noted certain quantitative discrepancies.
First the transitional range is shifted downwards somewhat more importantly than in the DNSs~(Manneville \& Rolland 2011).
This can be understood by noticing that the amount of energy extracted from the base flow by viscous stresses at the plates is transferred though a very short range of wall-normal small scales most likely to dissipate it.
This artificially maintains more turbulent activity in the system at given $R$, or equivalently a similar turbulence level at lower $R$, than in better resolved simulations and laboratory experiments where energy is transferred to and efficiently dissipated in much smaller scales.
(There is little or no trade-off for the in-plane dissipation that is treated like in the full-3D DNSs.)
For a concrete comparison, experiments (Prigent {\it et al.\/} 2002) and well-resolved full-3D DNSs (Duguet {\it et al.} 2010) give the upper threshold (featureless turbulence) at $R_{\rm t}\approx 410$ and the lower threshold (global stability) at $R_{\rm g}\approx 325$.
In simulations at reduced resolution (Manneville \& Rolland 2011), $N_y$ being the number of Chebyshev polynomials used in the representation of the wall-normal dependence, these values are shifted down to 350 and 270 for $N_y=15$ and to 275 and 215 for $N_y=11$.
Here we have $R_{\rm t} \approx 159$ and $R_{\rm g} \approx 150$ but the pattern is still well rendered.
This larger shift can therefore be understood because the effective wall-normal resolution is much lower.
The fact that  a physically relevant solution is obtained here while the  Chebyshev implementation breaks down with similarly few modes is due the optimal rendering of boundary conditions on $v$ achieved by our basis choice (see below).

Second, though the angle between the bands and the streamwise direction is correct, the wavelengths of the pattern, both streamwise and spanwise, are too short by a factor of 1.5 to 2, and the pattern's orientation in domains of intermediate size fluctuates more than in the DNSs.
The amount of enhancement is however difficult to appreciate quantitatively.
These phenomena remain unexplained for the moment but might relate to the effect of the wall-normal resolution on the streamwise coherence that was shown to play an important role on the existence of the pattern (Philip \& Manneville 2011).
A~hand-waving confirmation of this effect on the robustness of the bands comes from the continuous trend observed as the resolution is decreased, here as the truncation level $i_{\rm max}$ is lowered, in rough correspondence with what was observed when reducing the wall-normal resolution in DNSs.
First, a conspicuous steady pattern is observed  with $i_{\rm max}=3$.
Next, for $i_{\rm max}=2$ (not reported here but studied in parallel) coexistence of fluctuating, wide, laminar and turbulent domains are observed in an even narrower Reynolds range; these domains remain disorganised and do not form bands.
Finally, for $i_{\rm max}=1$~(Lagha \& Manneville 2007a), streaks stay short, the transitional range seems to be reduced to a point at a somewhat larger value ($R_{\rm g}\approx170$), and wide steady coexisting  domains do not exist: either turbulent domains grow from small germs in a laminar background or the reverse (Manneville 2009).
As a matter of fact, a similar but worse situation occurred in full 3D simulations at exaggeratedly reduced resolution since nontrivial states with unphysical small scales and no patterns were obtained for $N_y=9$ and $7$ and blow-up occurred for $N_y < 7$, which give a marked advantage to our separately optimal wall-normal representations of $v$ and $\zeta$.
These observations should contain some physics that warrant to be elucidated, as suggested in the next section.

\section{Perspectives\label{S4}}

Understanding the transition to turbulence in wall-bounded flows, and especially PCF that is linearly stable for all $R$ and displays alternating laminar and turbulent oblique stripes on its way to fully developed turbulence, is a hard problem when starting from the NSE.
Some simplification can be expected by taking a key ingredient into account: the transition takes place at moderate values of the Reynolds number for which the flow is controlled by the presence of coherent structures~(Hof {\it al al.} 2004; Bottin {\it et al.} 1998).
The model that we have derived incorporates this feature by means of a Galerkin expansion of the dynamics optimally adapted to the boundary conditions at the walls.
Truncating the expansion beyond lowest nontrivial order, keeping 7~amplitudes instead of 3 in (Lagha \& Manneville 2007a,b), has allowed us to recover the experimentally and numerically observed patterning at minimal price (downward shift of the transitional range, somewhat too short pattern wavelengths).
This limitation can be overcome by increasing $i_{\rm max}$, which raises the interesting question of the rate of convergence of the approximation.
Such a study would possibly be rewarding because, to be precise, DNSs treating the three space directions on a similar footing are computationally extremely demanding~(Duguet {\it et al.} 2010).
In a spirit akin to that of large eddy simulations, the present modelling avoids to waste numerical resources by singling out hydrodynamic coherence in the wall-normal direction and skipping the explicit computation of the small scales.

An indication that the convergence of our approach could be fast is the observation reported by one of us -- see Fig.~B6 in (Manneville 2015) -- that the fraction of the perturbation energy retained in projections of fully resolved numerical  solutions onto the basis considered here tends to 1 exponentially fast as the truncation order is increased: $90\%$, $97\%$, and $99\%$ for $i_{\rm max}=1$, $3$, and $5$, respectively.
This does not prove that the global dynamics of the system would be equally well captured quantitatively by increasing the number $2i_{\rm max}+ 1$ of fields in the model but hints at such a convergence, as generally expected for spectral approaches, here relying on specific complete series of Jacobi polynomials (Rolland 2012).
At the price of a pre-treatment of the problem that amounts to the once-for-all automated derivation of an effective set of equations of sufficiently high order, it might be found interesting to replace the full 3 D numerical simulations of the NSE by a finite set of two-dimensional partial differential equations already taking the continuity condition fully into account and   managing with wall-normal coherence in the transitional range.
A quantitative estimate of the expected gain in terms of memory requirements and time steps definitely warrants further study. 

In a complementary perspective, one can rather think of analysing the properties of the model.
First,  in-plane coherence may be added to the wall-normal coherence inherent in the derivation.
This can be done by inserting specific assumptions about the $(x,z)$-dependence of fields $\phi$ and $\psi$, in particular strict periodicity in space at the scale of the MFU~(Jim\'enez \& Moin 1991).
With $i_{\rm max} =1$  and further limiting the in-plane expansion to the first harmonic, it is then straightforward to recover Waleffe's models~(Waleffe 1997) by making the corresponding educated guess.
A system of eight equations for eight amplitudes is obtained, identical to his system~(10) but with a different set of coefficients acknowledging  the difference in boundary conditions (which, in passing, shows the structural genericity of that model).
For example,  the equation for the streamwise mean-flow component called $M$ by Waleffe and governed by his equation (10a) here reads:
$$
\label{wa}
\sfrac{\rm d}{{\rm d}t} \overline U_1 - \nu \bar p_{11} \overline U_1 = \sfrac14 \gamma \bar s_{101}\left[ (\alpha^2+\gamma^2) BE - 2 UV\right]
$$
where $\overline U_1\equiv M-1$, while other symbols have the same definition as in~(Waleffe 1997), especially the streamwise and spanwise wave-vectors  $\alpha=2\pi/\ell_x$ and $\gamma=2\pi/\ell_z$, $\ell_x$ and $\ell_z$ being the dimensions of the MFU.
The numerical values of coefficients in the equation above can be obtained from the formal expressions in~\ref{A}.
Following the very same line, a study  to be presented elsewhere~(Manneville, in preparation) shows that uniform large scale flows are generated just by shifting the phase of specific ingredients of Waleffe's eight-equation model.
Combining this to the introduction of appropriately weighted in-plane second harmonics should help us to account for oblique coherent structures like those recently found by Daly and Schneider~(2014), though the actual derivation of a model possessing them as fixed points would be cumbersome.

Beyond the simple hypotheses corresponding to strictly periodic coherent structures, the next step is to describe spatially slow turbulence modulations corresponding to the patterns observed experimentally through the formal introduction of a slow dependence of the amplitude of the local bifurcated state, in the spirit of the derivation of standard multiple-scale envelope equations.
The approach cannot be made as rigorous as, e.g., for convection the since the bifurcated state stays at finite distance from the laminar-flow base, which leaves room for further modelling.
Of the two scales introduced, the fast one accounts for mechanisms at the MFU scale and the slow one corresponds to the modulations.
The slow variables are driven by source terms arising from a filtering of the Reynolds stresses, like in~(Lagha \& Manneville 2007b) and it can be seen that the modulation of the uniform large scale flows alluded to above generates nonlocal contributions of the class identified by Hayot and Pomeau~(1994) as playing an essential role in the balance between laminar and turbulent regions responsible for patterning.
But, in contrast with their phenomenological introduction of such contributions, here they directly arise from the equations and are therefore sensitive to the local orientation of the flow with respect to the streamwise direction, hopefully giving a microscopic support to the empirical observations of Duguet and Schlatter (2013).

Finally, large scale flows are present already with $i_{\rm max} = 1$~(Lagha \& Manneville 2007b; Manneville, in preparation), though steady patterns are not observed in that case~(Manneville 2009).
Taking smaller wall-normal scales into account ($i_{\rm max} >1$) is therefore necessary for a theoretical interpretation of the stabilisation of long-wave modulations observed with $i_{\rm max}=3$, as reported in \S\ref{S3}.
Simplification of models with higher truncation levels would  then take advantage of the diagonal dominance of 
matrices $\mathbb A$, $\mathbb P$, and $\bar \mathbb P$ noticed earlier to perform the adiabatic elimination of terms of least relevance yielding an effective model for the slowly evolving terms.
Such a heavy work could however possibly not be necessary and considering seven fields might be sufficient up to an optimisation of the model's coefficients.
As a matter of fact, the three first amplitudes $(\Psi_0, \Psi_1,\Phi_1)$ are the most appropriate to deal with the nontrivial properties of the in-plane flow dependence.
So, if one is willing to include more of the wall-normal dependence, it should suffice to consider that the pairs $(\Psi_2,\Phi_2)$ and $(\Psi_3,\Phi_3)$ collect all the higher order contributions of each parity and, owing to its generic structure, to restrict oneself to the consideration of the seven-field model as an effective system replacing the NSE.
In this perspective, except as a starting guess, sticking to the values of the coefficients obtained in the strict Galerkin expansion is not advisable and introducing some multiplicative randomness at appropriate strategic places like in (Barkley 2011b) seems profitable.
Applying the program sketched in the previous paragraph to this new primitive problem is currently developed, which is expected to improve over the one-dimensional phenomenological approaches of Manneville (2012) and~Hayot \& Pomeau (1994).

\section{Conclusion}

The subcritical coexistence of different regimes forming laminar and turbulent patterns in PCF and other wall-bounded flow configurations is a difficult problem in which the interplay of mean flow corrections and finite amplitude perturbations plays a crucial role. 
Our approach {\it via\/}~Galerkin decomposition yields explicit models replacing the NSE by coupled systems governing amplitudes that encode the gross features of the flow.
The derivation is systematic and the structure of the obtained models is generic, reflecting that of the primitive equations.
Simulations of those models reproduce the patterning provided that the truncation level is not too low. 
They are next amenable to further analysis, especially through in-plane space dependence assumptions and explicit scale separation.
Here, this program has been developed for PCF but its adaptation to other flows such as plane Poiseuille or Couette--Poiseuille flow, cylindrical Couette--Taylor flow, etc. is straightforward.
Obtaining a Barkley-like model for Hagen--Poiseuille flow from first principles can also be considered along similar lines, using basis functions adapted to the tube geometry and the no-slip condition.
The extension to the less trivial case of boundary layer flows of various kinds, with their free-stream boundaries at infinity, remains a stimulating challenge.
Once obtained  such models offer tools to scrutinise laminar-turbulent coexistence and provide us with detailed physical explanations of this phenomenon of great conceptual and practical importance.

\appendix
\section{Explicit expressions\label{A}}
\subsection{Basis functions\label{Af}}
\noindent $\bullet$ In-plane velocity components and wall-normal vorticity component:
\begin{eqnarray*}
f(0)&=& \mbox{$1\over4$}\sqrt{15}\left(1-y^2\right),\\
f(1)&=& \mbox{$1\over4$}\sqrt{105}\left(1-y^2\right) y\,,\\
f(2)&=& \mbox{$21\over8$}\sqrt{5}\left(1-y^2\right) \left(y^2-\mbox{$1\over7$}\right),\\
f(3)&=& \mbox{$3\over8$}\sqrt{1155}\left(1-y^2\right) \left(y^2-\mbox{$1\over3$}\right) y\,,\\
f(4)&=&  \sfrac{33}{64}\sqrt{2730}\left(1- y^2\right) \left( y^4-\sfrac{6}{11}y^2 + \sfrac{1}{33}\right),\\
f(5)&=& \sfrac{429}{64}\sqrt{70}\left(1-y^2\right) \left(y^4 - \sfrac{10}{13}y^2 +\sfrac{15}{143}\right) y\, ,\dots
\end{eqnarray*}
\noindent $\bullet$ Wall-normal velocity component:
\begin{eqnarray*}
g(1)&=& \sfrac3{16}\sqrt{35}\left(1-y^2\right)^2,\\
g(2)&=&\sfrac3{16}\sqrt{385}\left(1- y^2\right)^2 y\,,\\
g(3)&=&\sfrac{33}{32}\sqrt{91}\left(1-y^2\right)^2\left(y^2- \mbox{$1\over{11}$}\right),\\
g(4)&=& \sfrac{39}{32}\sqrt{385} \left(y^2 - 1\right)^2\left(y^2- \sfrac{3}{13}\right) y\,,\\
g(5)&=& \sfrac{195}{128}\sqrt{1309}\left(y^2 - 1\right)^2\left(y^4 - \sfrac{2}{5}y^2 +\sfrac{1}{65}\right)\,, \dots
\end{eqnarray*}

\subsection{Coefficients in the evolution equations\label{Ag}}
Taking care of the order of the subscripts introduced, symmetries within the sets of coefficients are easily detected, directly or {\it via\/} integration by parts. Energy conservation relies on the symmetries of coefficients introduced in the expressions of nonlinear terms.
The elements of matrix $\mathbb C$ appearing in (\ref{eq:UW}) are straightforwardly obtained as \quad $c_{ji}=\int_{-1}^{+1}\DY f_j g'_i\,$.

\subsubsection{Equation (\ref{M1}) for $\Phi_j$}\mbox{}

\noindent $\bullet$ linear terms:\\
 \hspace*{1em}matrix $\bar\mathbb I$:\quad $\bar\delta_{ji}=\int_{-1}^{+1}\DY g_j(y g_i)$,\\
 \hspace*{1em}matrices $\mathbb A$ and $\bar \mathbb A$:\quad $a_{ji}=\int_{-1}^{+1}\DY  g_j g''_i$,\quad
   $\bar a_{ji}=\int_{-1}^{+1}\DY g_j (y g''_i)$,\\
\hspace*{1em}matrix $\mathbb P$:\quad $p_{ji}=\int_{-1}^{+1}\DY g_j g''''_i$;\\[2ex]
$\bullet$ nonlinear terms, for $j\in(1:i_{\rm max})$:
 \begin{eqnarray}
\nonumber N^{(V)}_j&=&\sum_{i=0}^{i_{\rm max}}\sum_{k=1}^{i_{\rm max}}q_{jik}  \Delta\left[\partial_x(U_iV_k)+\partial_z(W_iV_k)\right] +\sum_{i=1}^{i_{\rm max}}\sum_{k=1}^{i_{\rm max}}\bar q_{jik}\Delta (V_i V_k)\\
\nonumber&&\mbox{}-\sum_{i=0}^{i_{\rm max}}\sum_{k=0}^{i_{\rm max}}\,r_{jik}\left[\partial_{xx} \left(U_iU_k\right) + 2\partial_{xz}\left(U_iW_k\right)+\partial_{zz} \left(W_iW_k\right)\right]\\
\label{e3}&&\mbox{}-\sum_{i=0}^{i_{\rm max}}\sum_{k=1}^{i_{\rm max}}\bar r_{jik}\left[\partial_x\left(U_iV_k\right)+\partial_z\left(W_iV_k\right)\right],
\end{eqnarray}
\hspace*{1em}with:
$q_{jik}=\int_{-1}^{+1}\DY g_j f_i g_k$, \quad $\bar q_{jik}=\int_{-1}^{+1}\DY g_j (g_i g_k)'$,\\
\hspace*{1em}
$r_{jik}= \int_{-1}^{+1}\DY  g_j(f_if_k)'$,\quad $\bar r_{jik}=\int_{-1}^{+1}\DY  g_j(f_ig_k)''$.

\subsubsection{Equation (\ref{M2}) for $\Psi_j$}\mbox{}

\noindent $\bullet$  linear terms:\quad matrices $\mathbb B$, $\bar\mathbb B$, and $\bar\mathbb P$:\\
 \hspace*{1em}$b_{ji}=\int_{-1}^{+1}\DY f_j (yf_i)$,\quad $\bar b_{ji}=\int_{-1}^{+1}\DY f_j g_i$,\quad $\bar p_{ji}=\int_{-1}^{+1}\DY f_j f''_i$,\\[2ex]
$\bullet$ nonlinear terms, for $j\in(0:i_{\rm max})$:
\begin{eqnarray}
\nonumber N^{(Z)}_j&=&\sum_{i=0}^{i_{\rm max}}\sum_{k=0}^{i_{\rm max}}s_{jik}\left[\partial_{xz} (U_iU_k - W_iW_k) + (\partial_{zz}-\partial_{xx})(U_iW_k)\right]\\
\label{e4}&&\mbox{}+\sum_{i=0}^{i_{\rm max}}\sum_{k=1}^{i_{\rm max}}\bar s_{jik} \left[\partial_z(U_iV_k) -\partial_x(W_iV_k))\right]\!,
\end{eqnarray}
\hspace*{1em}with
$s_{jik}=\int_{-1}^{+1}\DY f_j f_i f_k,\quad \bar s_{jik}=\int_{-1}^{+1}\DY f_j (f_i g_k)'$.

\paragraph{Acknowledgements.} Results described here have been obtained by K.S. within the framework of program ``Fluid Mechanics, Fundamental \& Applications'' of \'Ecole Polytechnique's Master of Mechanics under the supervision of P.M.
Thanks are due to Y. Duguet (LIMSI, Orsay, France) for interesting discussions about the topics treated here and his critical reading of the manuscript.
Constructive remarks of the Referees are also deeply acknowledged. 

\section*{References}

\parindent 0pt
\parskip 1ex
Barkley D  2011
(a) Simplifying the complexity of pipe flow, {\em Phys. Rev. E} {\bf 84},~016309;
(b) Modeling the transition to turbulence in shear flows,
{\em J. Phys.: Conf. Ser.} {\bf 318}, 032001.

Barkley D  and  Tuckerman L  2005 
Computational study of turbulent laminar patterns in Couette flow,
{\em Phys. Rev. Lett.} {\bf 94},~014502.

Bottin S, Dauchot O, Daviaud F  and  Manneville P  1998
Experimental evidence of streamwise vortices as finite amplitude solutions in transitional plane Couette flow,
{\em Phys. Fluids} {\bf 10},~2597--2607.

Canuto C, Hussaini M,  Quarteroni A and Zang T  2007 {\em Spectral methods:
  evolution to complex geometries and applications to fluid dynamics}, Springer.

Daly C  and  Schneider T 2014
Oblique coherent structures in plane {Couette} flow,
in {\em Subcritical transition to turbulence, Euromech EC565 Colloquium\/} edited by  Duguet Y, Wesfreid J and  Hof~B.

Duguet Y  and  Schlatter P  2013
Oblique laminar-turbulent interfaces in plane shear flows,
{\em Phys. Rev. Lett.} {\bf
  110},~034502.

Duguet Y, Schlatter P  and  Henningson D  2010
Formation of turbulent patterns near the onset of transition in plane Couette flow,
{\em J. Fluid Mech.} {\bf  650}, 119--129.

Finlayson B  1972 {\em The Method of Weighted Residuals and Variational Principles}, Academic Press.

Hayot F  and  Pomeau Y  1994
{T}urbulent domain stabilization in annular flows,
{\em Phys. Rev. E} {\bf 50},~2019--2021.

Hof B, van Doorne C, Westerweel J, Nieuwstadt F, Faisst H, Eckhardt B, Wedin H, Kerswell R  and  Waleffe F  2004
 Experimental observation of nonlinear traveling waves in turbulent pipe flow,
  {\em Science} {\bf 305},~1594--1598.

Huerre P  and  Rossi M  1998 Hydrodynamic instabilities in open flows, in {\em Hydrodynamics and Nonlinear Instabilities}, edited by Godr\`eche C and Manneville P Cambridge University Press, pp.~81--294.

Jim\'enez J  and  Moin P  1991
The minimal flow unit in near wall turbulence,
{\em J.~Fluid Mech.} {\bf 225},~213--240.

Lagha M  and  Manneville P  2007 (a) Modeling transitional plane Couette flow, {\em Eur. Phys. J. B} {\bf 58},~433--447. (b) Large scale flow around turbulent spots, {\em Phys. Fluids} {\bf 19},~094105.

Manneville P  2009
Spatiotemporal perspective on the decay of turbulence in wall-bounded flows,
{\em Phys. Rev. E} {\bf 79}, 025301 (R).

Manneville P  2012
Turbulent patterns in wall-bounded flows: A {T}uring instability?,
{\em Europhys. Lett.} {\bf 98},~64001.

Manneville P  2015
On the transition to turbulence of wall-bounded flows in general, and plane {C}ouette flow in particular,
 {\em Eur. J. Mech. B-Fluids} {\bf49}, 345--362.

Manneville P  On the generation of large scale flows in transitional wall-bounded flows,
\newblock In preparation.

Manneville P  and  Rolland J  2011
On modelling transitional turbulent flows using under-resolved direct numerical simulations: the case of plane Couette flow,
{\em Theor. Comp. Fluid Dyn.} {\bf 25},~407--420.

Philip J  and  Manneville P  2011
From temporal to spatiotemporal dynamics in transitional plane {C}ouette flow,
{\em Phys. Rev. E} {\bf 83},~036308.

Prigent A, Gr\'egoire G, Chat\'e H, Dauchot O  and  van Saarloos W 2002
Large-scale finite-wavelength modulation within turbulent shear flows,
 {\em Phys. Rev. Lett.} {\bf 89},~014501.

Rolland J 2012
\'Etude num\'erique \`a petite et grande \'echelle de la bande laminaire-turbulente de l'\'ecoulement de Couette plan transitionnel,
{\em PhD thesis, \'Ecole Polytechnique\/} (in French)\\
{\tt https://tel.archives-ouvertes.fr/pastel-00755414}

Rolland J  and  Manneville P  2011
Pattern fluctuations in transitional plane Couette flow,
{\em J. Stat. Phys.} {\bf 142},~577--591.

Schmid P  and  Henningson D  2001 {\em Stability and Transition in Shear Flows}, Springer.

Waleffe F 1997
On a self-sustaining process in shear flows,
{\em Phys. Fluids} {\bf 9},~883--900.

\end{document}